\documentclass[aps,pra,numerical,reprint,showpacs]{revtex4-1}  
\usepackage[dvips]{graphicx}%
\usepackage{bm}%
\usepackage{amsmath}
\usepackage{mathrsfs}
\usepackage{dcolumn}
\usepackage[normalem]{ulem} 
\usepackage[colorlinks=true,linkcolor=blue,citecolor=blue ,urlcolor=blue]{hyperref}

\begin{document}

\newcommand{\op}[1]{{\bm{#1}}}
\newcommand{\bra}{\langle}
\newcommand{\ket}{\rangle}
\newcommand{\new}[1]{\textcolor[rgb]{0,0.3,0}{\uline{#1}}}
\newcommand{\old}[1]{\textcolor[rgb]{1,0,0}{\sout{#1}}}
\newcommand{\nota}[1]{\textcolor[rgb]{0,0.5,0.5}{{#1}}}


\title{Variational Study of $\lambda$- and $N$-Atomic Configurations Interacting with an Electromagnetic Field of $2$ Modes}
\author{S. Cordero} 
\email{sergio.cordero@nucleares.unam.mx}
\author{O. Casta\~nos}
%
\author{R. L\'opez-Pe\~na}
%
\author{E. Nahmad-Achar}

%
\affiliation{%
Instituto de Ciencias Nucleares, Universidad Nacional Aut\'onoma de M\'exico, Apartado Postal 70-543, 04510 Mexico City CDMX, Mexico }

\date{}

\begin{abstract}

A study of the $\lambda$- and $N$-atomic configurations under dipolar interaction with $2$ modes of electromagnetic radiation is presented. The corresponding quantum phase diagrams are obtained by means of a variational procedure. Both configurations exhibit normal and collective (super-radiant) regimes. While the latter in the $\lambda$-configuration divides itself into $2$ subregions, corresponding to each of the modes, that in the $N$-configuration may be divided into $2$ or $3$ subregions depending on whether the field modes divide the atomic system into $2$ separate subsystems or not.

Our variational procedure compares well with the exact quantum solution. The properties of the relevant field and matter observables are obtained.

\end{abstract}


\maketitle

\section{Introduction}

The interaction between matter and radiation has been determinant in the study of quantum optics and information science. A simple and useful model was proposed by Dicke~\cite{dicke54}, which enhances the cooperative nature of the spontaneous emission from a system of identical atoms. In this model the system suffers a transition from a normal to a superradiant phase~\cite{hepp73, hepp73b}. Artificial atoms can be realized as an approximation to real atoms~\cite{kastner93,astafiev10,buluta11}, which exhibits the transition to this superradiant regime~\cite{baumann10, baksic13}.

A first generalization of the model is the consideration of atoms of three or more levels~\cite{yoo85,abdel-wahab07,abdel-wahab08}. These allow to consider dipolar interactions with one, two, or more modes of electromagnetic field, and have been extensively studied~\cite{yoo85,civitarese1,abdel-wahab07,cordero1,cordero2, hayn11,hayn12, cordero15}.

Recently, the generalized Dicke model has been studied to determine the quantum phase diagrams of $N_a$ atoms of $n$-levels interacting dipolarly with $\ell$ modes of electromagnetic field, where each mode promotes transitions only between two given atomic levels~\cite{cordero15}. In this work we generalize this study to the case where one mode can connect more than one pair of levels in the particular $4$-level atomic configurations $\lambda$ and $N$.  

The quantum phase diagrams are obtained by means of a variational procedure, which allows for analytical expressions both for the critical variables and the expectation values of the field and matter operators, and confirmed through the exact quantum calculation (even for a value of $N_a$ as low as $1$).

This paper is organized as follows: 
Section~\ref{model} describes the Hamiltonian for a system of $N_a$ atoms in the $4$-level $\lambda$ and $N$ configurations, under dipolar interaction with $2$ modes of electromagnetic field. A transition between a given pair of levels is promoted by only one mode, but one mode may promote transitions between more than one pair of levels.
A variational test function used to calculate the energy surface of the $\lambda$ system is given in section~\ref{s.lambda}, and the ground state is estimated by minimizing it with respect to the field and matter variables. This allows for analytical expressions for the expectation values of the relevant field and matter operators, and the results are compared with the exact quantum solution.
The corresponding analysis for the $N$-configuration is presented in section~\ref{s.N}.
Section~\ref{conclusions} gives some concluding remarks.

\section{Model}
\label{model}

The full Hamiltonian under the dipolar approximation consists of a diagonal and an interaction contribution. The first is given by

\begin{equation}
	\op{H}_D = \Omega_1\,\op{a}_1^\dag\,\op{a}_1 + \Omega_2\,\op{a}_2^\dag\,\op{a}_2 + \sum_{k=1}^4 \omega_k\,\op{A}_{kk}\,,
\end{equation}
where $\op{a}^\dagger_s,\ \op{a}_s$ ($s=1,2$) are the creation and annihilation operators for the mode $s$; $\Omega_s$ and $\omega_k$ for $s=1,\,2$ and $k=1,\,\dots,\,4$ are respectively the field and the atomic level frequencies, and we have adopted the convention $\omega_1 \leq \omega_2 \leq \omega_3 \leq \omega_4$. The matter operators $\op{A}_{jk}$ are the generators of $U(4)$, and for identical particles they may be represented with a bosonic realization $\op{A}_{jk} = \op{b}_j^\dag\,\op{b}_k$.
The first order Casimir operator $\sum_{k=1}^4 \op{A}_{kk} = N_a \, \op{I}$ represents the total number of atoms in the system, which is a constant of motion.

\begin{figure}[h!]
	\begin{center}
\includegraphics[width=0.7\linewidth]{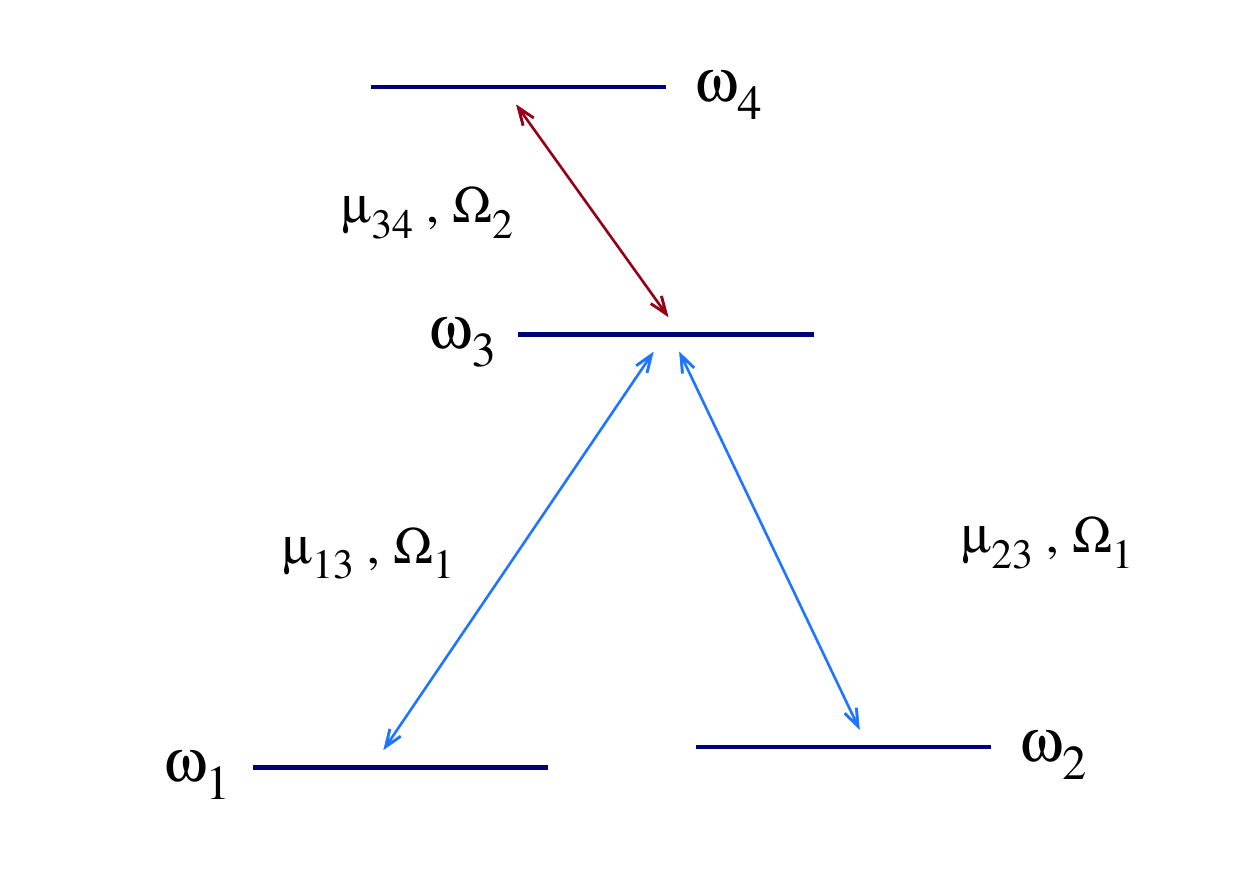} \\
\includegraphics[width=0.7\linewidth]{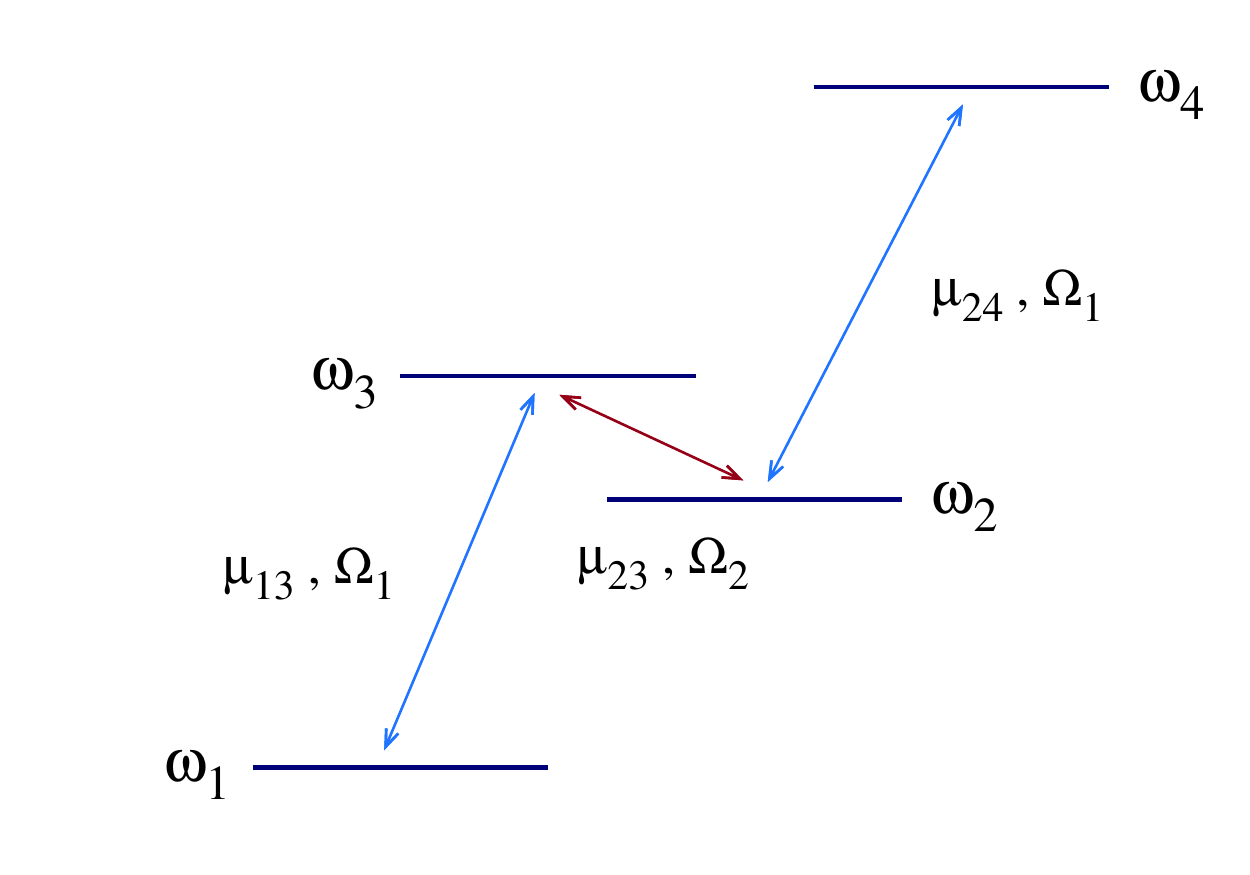}
	\end{center}
\caption{(color online) Schematic diagrams for the $\lambda$ (top) and $N$ (bottom) atomic configurations interacting with $2$ radiation modes. In each case the nonzero dipolar strengths $\mu_{jk}$ are indicated.}
\label{f.examples}
\end{figure}

For the interaction Hamiltonian we have, in general,
\begin{equation}\label{Hint}
\op{H}_{ij}^{(s)} = - \frac{1}{\sqrt{N_a}}\mu_{ij} \left(\op{A}_{ij}+\op{A}_{ji}\right)\left(\op{a}_s^\dag+\op{a}_s\right) \,,
\end{equation}
where the dipolar strength for the transition $\omega_i \rightleftharpoons \omega_j$ has been denoted by $\mu_{ij}$. The upper index $s$ indicates the mode of the electromagnetic field.

From Fig.\ref{f.examples} one determines the form of the Hamiltonians for each configuration, i.e.,
\begin{subequations}
	\begin{eqnarray}
	 	\label{Hlambda}
	\op{H}_\lambda &=& \op{H}_D + \op{H}_{13}^{(1)} + \op{H}_{23}^{(1)} + \op{H}_{34}^{(2)} \, ,  \\[3mm]
		\label{HN}
	\op{H}_N &=& \op{H}_D + \op{H}_{13}^{(1)} + \op{H}_{23}^{(2)} + \op{H}_{24}^{(1)}\,.
	\end{eqnarray}
\end{subequations}

\section{$\lambda$-configuration}
\label{s.lambda}

\subsection{Variational ground state}

In order to find the variational energy surface we use as a test function the direct product of coherent states for the field and matter. For the field contribution we take the Heisenberg-Weyl states $|\vec{\alpha}\ket = |\alpha_1\ket\otimes|\alpha_2\ket$, and for the matter contribution we take the totally symmetric $U(4)$ coherent states $|\vec{\gamma}\ket = |\gamma_1\,\gamma_2\,\gamma_3\,\gamma_4\ket$~\cite{lopez-pena15}, 
\begin{eqnarray}
|\vec{\alpha}\ket&=& \mathcal{C}_n \sum_{\nu_1, \nu_2} \frac{N^{\frac{\nu_1 +\nu_2}{2}}_a \,  e^{i (\theta_1 \nu_1 + \theta_2 \nu_2 )} \, r^{\nu_1}_1\, r^{\nu_2}_2 }{\nu_1! \, \nu_2!} \vert \nu_1 \nu_2 \rangle_F \, , \nonumber \\
\vert \vec{\gamma}\ket&=& \frac{1}{ \sqrt{N_a! \, \sum_{k=1}^4 \varrho_k^2}} \, \left( \sum_{k=1}^4 \varrho_k \, e^{i \phi_k} \, \op{b}^\dagger_k \right)^{N_a} \vert 0\rangle_M \, .
\label{varstate}
\end{eqnarray}
where $\mathcal{C}_n=e^{-N_a (r^{2}_1+ r^{2}_2)/2}$ is a normalization factor, $\op{b}_k \vert 0 \rangle=0$, $\alpha_s = \sqrt{N_a}\,r_s\,e^{i\theta_s}$, and $\gamma_k = \varrho_k\, e^{i\phi_k}$.

The variational energy surface per particle for the Hamiltonian~(\ref{Hlambda}) is given by
\begin{eqnarray}\label{eq.varE}
E&=& \Omega_1\, r_1^2+ \Omega_2\, r_2^2 + \frac{1}{\Gamma_0^2}\left( \omega_{1}+\omega_{2}\, \varrho_2^2 + \omega_{3}\, \varrho_3^2 +\omega_{4}\, \varrho_4^2 \right)\nonumber \\[3mm]
&-&\frac{4}{\Gamma_0^2}r_1\,\varrho_3\,\cos(\theta_1)\bigg( \mu_{13}\,\cos(\phi_{13})+ \mu_{23}\,\varrho_2\,\cos(\phi_{23})\bigg)\nonumber\\[3mm]
&-&\frac{4}{\Gamma_0^2}\mu_{34} \,r_2\,\varrho_3\,\varrho_4\,\cos(\theta_2)\,\cos(\phi_{34})\,,
\end{eqnarray}
with $\phi_{jk} = \phi_k-\phi_j,\ \Gamma_0^2 =\sum_{k=1}^4 |\gamma_k|^2$, and where we have substituted explicitly $\varrho_1 = 1$.

The critical values of the field variables may be obtained immediately,
\begin{eqnarray}
	\label{crit_ang}
\theta_i^c &=& 0,\,\pi,\qquad \phi_{ik}^c = 0,\,\pi,\,\\[3mm]
	\label{crit_r}
r_1^c &=& \frac{2\,\left(\mu_{13}+\mu_{23}\,\varrho_2^c\right)\, \varrho_3^c}{\Omega_1\,\Gamma_0^{c\,2}} \,,\quad
r_2^c = \frac{2\,\, \mu_{34}\,\varrho_3^c\, \varrho_4^c}{\Omega_2\,\Gamma_0^{c\,2}} \,,\qquad 
\end{eqnarray}
while those for the matter variables take different values depending on the region in parameter space.

\begin{figure}[h!]
\includegraphics[width=0.85\linewidth]{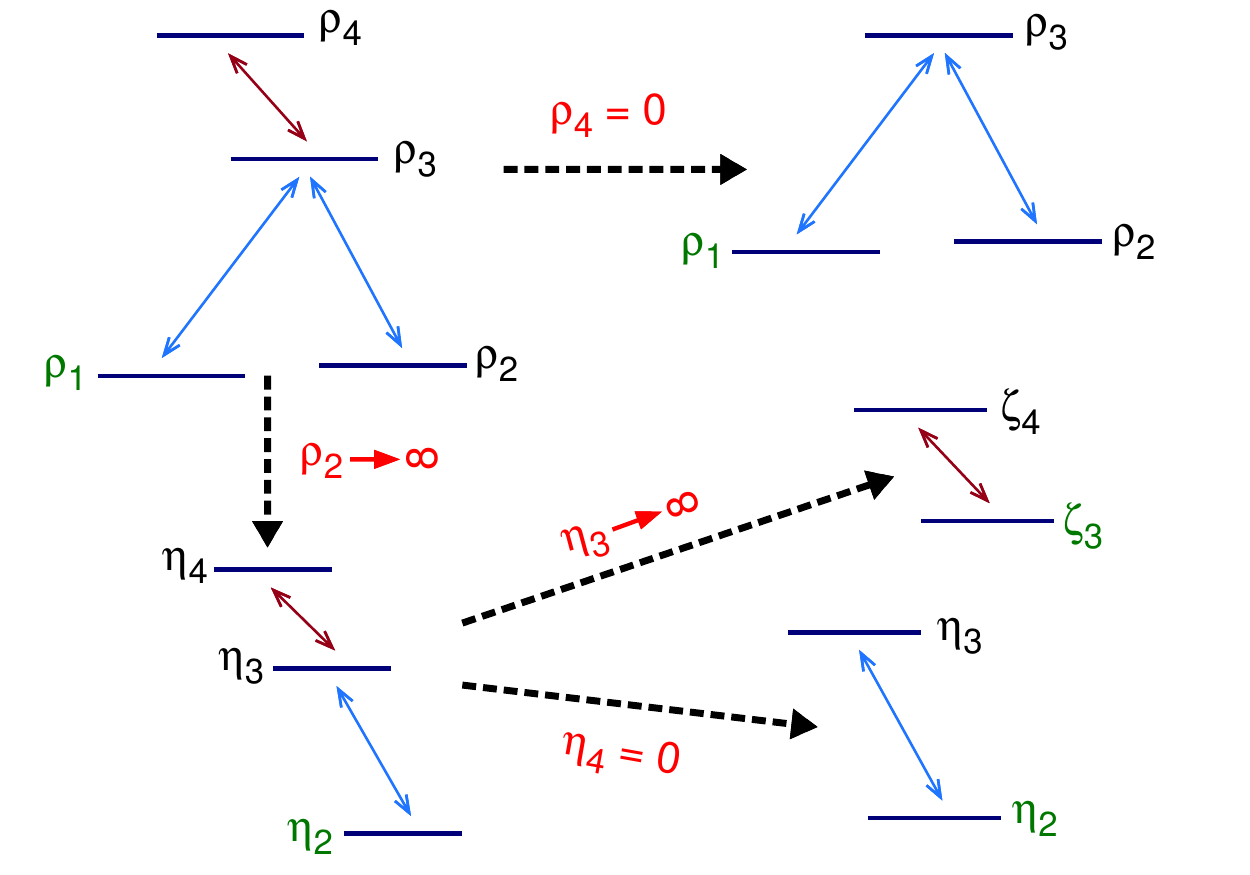}
\caption{(color online) Reduction of the $4$-level $\lambda$-configuration to $3$- and $2$-level configurations in the collective regime. See text for details.}
\label{lambda_reduction}
\end{figure}

The values $\varrho_2^c=\varrho_3^c=\varrho_4^c=0$ provide the simple solution corresponding to the vacuum state for the field contribution (cf. Eq.~(\ref{crit_r})) and all atoms in their lower state. Substituting these critical values in equation~(\ref{varstate}) one finds the state $\vert 0\rangle_F \otimes \vert N_a,0,0,0\rangle_M$ and the minimum variational energy is
\begin{equation}
E_{norm} = \omega_1\,,
\end{equation}
which corresponds to the so-called {\it normal region} of the configuration. All other extrema yield a reduction of the $\lambda$ configuration, as given in what follows:

\vspace{0.1in}
{\em Reduction of the $\lambda$ configuration}.- When at least one critical value is non-zero the system reduces to subsystems with a lower number of levels~\cite{cordero15}. This is schematically shown in Fig.~\ref{lambda_reduction}. When $\varrho_4^c=0$ and $\varrho_2^c,\ \varrho_3^c$ remain finite, the equations for the critical points reduce to those of the $3$-level $\Lambda$-configuration for $\omega_1,\,\omega_2$ and $\omega_3$, interacting dipolarly with one mode ($\Omega_1$), for which one may to obtain analytical solutions~\cite{cordero1,cordero2} under the equal detuning ($\omega_2=\omega_1$) condition. This system has in turn its {\em normal} and {\em collective} regions, separated by the curve $4(\mu_{13}^2 + \mu_{23}^2) - (\omega_3-\omega_1)\Omega_1 = 0$ in the $\mu$-parameter space. The minimum energy surface (for equal detuning) is respectively given by $E_{{\Lambda}_{norm}} = \omega_1$ and
\begin{eqnarray}
E_\Lambda &=& \omega_1 - \frac{(4(\mu_{13}^2+\mu_{23}^2)-(\omega_3-\omega_1)\Omega_1)^2}{16\,(\mu_{13}^2+\mu_{23}^2)\Omega_1}\,.
\end{eqnarray}
The state in the normal region is $\vert 0\rangle_F \otimes \vert N_a,0,0,0\rangle_M$, and in the collective region is obtained by substituting in Eq.~(\ref{varstate}) the critical values for $S_{\Lambda}$ given in Table~\ref{t.lambda}.
Further reductions of this configuration to $2$ levels give results which are contained in the minima of $\Lambda$ by restricting oneself to the axes $\mu_{13}=0$ or $\mu_{23}=0$.       

The limit $\varrho_2\to \infty$ with $\varrho_k^c = \eta_k\,\varrho_2^c$ for $k=3,\,4$ leads (cf. Fig.~\ref{lambda_reduction}) to a $3$-level $\Xi$-configuration in the parameters $\eta_k\ (k=2,3,4)$ interacting with a $2$-mode field. Following  the same procedure~\cite{cordero1,cordero2}, this in turn reduces to $2$-level systems:

For $\eta_4=0$, one gets a $2$-level system in the variables $\eta_2,\,\eta_3$, with its normal and collective regions separated by $4\,\mu_{23}^2-(\omega_3-\omega_2)\Omega_1 = 0$ and minimum energies $E_{{23}_{norm}}~=~ \omega_2$ and
\begin{equation}
E_{23} = \omega_2 - \frac{(4\,\mu_{23}^2-(\omega_3-\omega_2)\Omega_1)^2}{16\,\mu_{23}^2\,\Omega_1}\,.
\end{equation} 
The states are $\vert 0\rangle_F \otimes \vert 0,N_a,0,0\rangle_M$ for the normal region, and the corresponding to Eq.~(\ref{varstate}) when the critical values for $S_{23}$ in Table~\ref{t.lambda} are substituted.

On the other hand by taking the limit $\eta_3\to\infty$,  a $2$-level system in the variables $\zeta_3,\,\zeta_4$ is obtained, with its normal and collective regions separated by the Maxwell set $4\,\mu_{34}^2-(\omega_4-\omega_3)\Omega_2 = 0$ and minimum energies $E_{{34}_{norm}} = \omega_3$ and
\begin{equation}
E_{34} = \omega_3 - \frac{(4\,\mu_{34}^2-(\omega_4-\omega_3)\Omega_2)^2}{16\,\mu_{34}^2\,\Omega_2}\,,
\end{equation}
The states are $\vert 0\rangle_F \otimes \vert 0,0,N_a,0\rangle_M$ for the normal region, and the corresponding to Eq.~(\ref{varstate}) when the critical values for $S_{34}$ in Table~\ref{t.lambda} are substituted.
\vspace{0.1in}

{\em Minimum energy surface}.- The minimum variational energy surface as a function of the dipolar strengths is then 
\begin{equation}
E_{\lambda_{min}} = \min\{E_{norm},\,E_{\Lambda},\,E_{23},\,E_{34}\}\,,
\end{equation} 
with the critical values of the Hamiltonian variables in each region shown in Table~\ref{t.lambda}.

\begin{figure}[!h]
\includegraphics[width=0.85\linewidth]{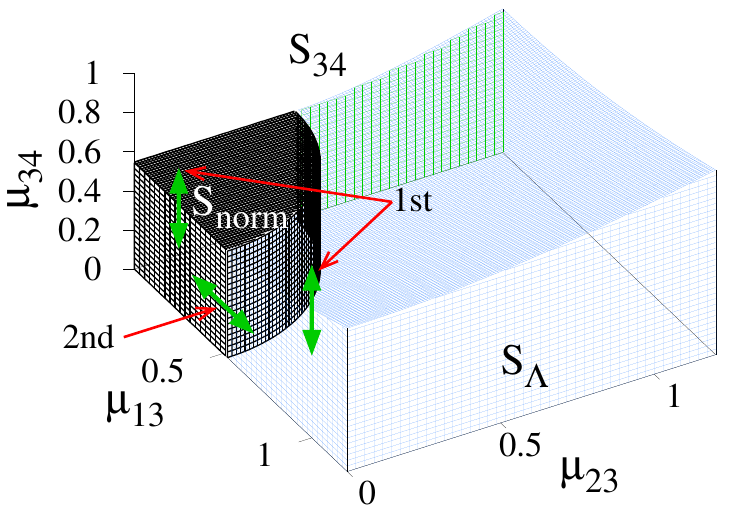}
\caption{(color online) Phase diagram of the $4$-level $\lambda$-configuration interacting dipolarly with two modes of electromagnetic field. The normal region is shown in black. The order of the transitions is also indicated. The collective region is divided into subregions where each mode dominates: $\Omega_1$ below the blue surface (subregion denoted by $S_\Lambda$) and $\Omega_2$ above it (subregion denoted by $S_{34}$). The region $S_{23}$ corresponds to the back wall (in green), given by $\mu_{13}=0$ within the $S_{\Lambda}$ region. The parameter space is drawn for $\Omega_1=1,\,\Omega_2=0.25,\,\omega_1=\omega_2=0,\,\omega_3= 1.1$ and $\omega_4=1.3$.}
\label{f.sep.l}
\end{figure}

This divides the whole parameter space $(\mu_{13},\, \mu_{23},\, \mu_{34})$ into monochromatic subregions $S_i$, each of which is dominated by a mode of the radiation field, in agreement with~\cite{cordero15}. This is shown in Figure~\ref{f.sep.l} for the case of equal detuning $\omega_1 = \omega_2$, where the subindex of $S$ denotes the region where the corresponding energy dominates. In this, the normal region is shown in black and the order of the transitions is also indicated. As known~\cite{lopez-pena15}, the $3$-level $\Lambda$-configuration in the equal detuning has a second-order transition when the state goes from the normal to collective region $S_{norm}\rightleftharpoons S_\Lambda^{0}$, which is related to the fact that the critical points in the separatrix for this case form bifurcations. In general, both first and second order transitions for $S_{norm}\rightleftharpoons S_\Lambda$ occur~\cite{cordero2}. A first order transition occurs both for $S_{norm}\rightleftharpoons S_{34}$ and for $S_\Lambda^{(0)}\rightleftharpoons S_{34}$ because the critical points along this separatrix form a Maxwell set~\cite{castanos14,cordero15}, and this result remains in the general case $S_\Lambda\rightleftharpoons S_{34}$.

\begin{table*}[!ht]
\caption{Critical values for the $\lambda$-configuration under equal detuning $\omega_1=\omega_2=0$. Here, $\omega_{kj}=\omega_k-\omega_j$. Note that the values for the $S_{23}$ region may be obtained from those for$S_{\Lambda}$ taking the limit $\mu_{13} \to 0$.}
\label{t.lambda}
\vspace{2mm}
\begin{tabular}{l | c c c c c}
& $r_1^c$& $r_2^c$& $\varrho_2^c$ & $\varrho_3^c$ & $\varrho_4^c$ \\[1mm] \hline \hline  \\
$S_{norm}$ &0&0& 0&0&0 \\[2mm]
$S_\Lambda$ & $\displaystyle \frac{\sqrt{16\, \left(\mu_{13}^2+\mu_{23}^2\right)^2- \omega_{13}^2\,\Omega_1^2}}{4\,\sqrt{\mu_{13}^2+\mu_{23}^2}\,\Omega_1}$& 0 & $\displaystyle \frac{\mu_{23}}{\mu_{13}} $ & $\displaystyle  \sqrt{\frac{\left(\mu_{13}^2+\mu_{23}^2\right)\left(\mu_{13}^2+\mu_{23}^2-\omega_{31}\,\Omega_1\right)}{\mu_{13}^2\,\left(\mu_{13}^2+\mu_{23}^2+\omega_{31}\,\Omega_1\right)}}$ & 0 \\[2mm]
&&&&& \\
\hline&& &&& \\[-3mm] &&& $\varrho_2^c\to \infty$ &$\eta_3^c$& $\eta_4^c$\\[1mm]
\hline &&&&& \\[-2mm]
$S_{23}$&$\displaystyle \frac{\sqrt{16 \mu_{23}^4-\omega_{32}^2\,\Omega_1^2}}{4\,\mu_{23}\,\Omega_1}$&0&&$\displaystyle \sqrt{\frac{4\,\mu_{23}^2- \omega_{32}\,\Omega_1}{4\,\mu_{23}^2+ \omega_{32}\,\Omega_1}}$&0\\[2mm]
&&&&& \\
\hline&& &&& \\[-3mm] &&&  &$\eta_3^c\to\infty$ & $\zeta_4^c$\\[1mm]
\hline &&&&& \\[-2mm]
$S_{34}$ & 0& $\displaystyle \frac{\sqrt{16 \mu_{34}^4 - \omega_{43}^2\,\Omega_2^2}}{4\,\mu_{34}\,\Omega_2}$ & & & $\displaystyle \sqrt{\frac{4\,\mu_{34}^2 - \omega_{43}\,\Omega_2}{4\,\mu_{34}^2 + \omega_{43}\,\Omega_2}}$\\[3mm]
\end{tabular}
\end{table*}

The relevant expectation values of the field and matter observables may be calculated from these variational states analytically. These are shown in Table~\ref{t.lambda.ev}, where $\bra \op{\nu}_i\ket$ denotes the expectation value of the number of photons of mode $\Omega_i$ for $i=1,2$, and $\bra \op{A}_{kk}\ket$ that of the population of level $k$, for $k=1,\dots,4$.

%
\begin{table*}
\caption{Expectation values per particle for the different regions in parameter space in the $\lambda$ configuration; case of equal detuning $\omega_1=\omega_2=0$. Here, $\omega_{kj}=\omega_k-\omega_j$.}
\label{t.lambda.ev}
\vspace{2mm}
\begin{tabular}{c| c | c | c |c}
& $S_{norm}$& $S_\Lambda$&$S_{23}$& $S_{34}$  \\[1mm] \hline \hline  &&& \\
$\bra \op{H}_\lambda\ket$ &0&$- \displaystyle\frac{(4\,(\mu_{13}^2+\mu_{23}^2)-\omega_{31}\,\Omega_1)^2}{16\,(\mu_{13}^2+\mu_{23}^2)\,\Omega_1}$& $ - \displaystyle\frac{(4\,\mu_{23}^2-\omega_{32}\,\Omega_1)^2}{16\,\mu_{23}^2\,\Omega_1}$ &  $\displaystyle \omega_3 - \frac{(4\,\mu_{34}^2-\omega_{43}\,\Omega_2)^2}{16\,\mu_{34}^2\,\Omega_2}$ \\[5mm]
$\bra \op{\nu}_1\ket$ &0&$\displaystyle \frac{16 \,\left(\mu_{13}^2+\mu_{23}^2\right)^2- \omega_{31}^2\,\Omega_1^2}{16\,\left(\mu_{13}^2+\mu_{23}^2\right)\,\Omega_1^2}$& $\displaystyle \frac{16 \,\mu_{23}^4- \omega_{32}^2\,\Omega_1^2}{16\,\mu_{23}^2\,\Omega_1^2}$ & 0 \\[5mm]
$\bra \op{\nu}_2\ket$ &0&0&0& $\displaystyle \frac{16\, \mu_{34}^4 - \omega_{43}^2\,\Omega_2^2}{16\,\mu_{34}^2\,\Omega_2^2}$ \\[5mm]
$\bra \op{A}_{11}\ket$ &1&$\displaystyle \frac{\mu_{13}^2\,\left(4\,(\mu_{13}^2+\mu_{23}^2) + \omega_{31}\,\Omega_1\right)}{8\, (\mu_{13}^2+\mu_{23}^2)^2}$& 0 & 0 \\[5mm]
$\bra \op{A}_{22}\ket$ &0&$\displaystyle \frac{\mu_{23}^2\left(4\,(\mu_{13}^2+\mu_{23}^2) + \omega_{31}\,\Omega_1\right)}{8 \,(\mu_{13}^2+\mu_{23}^2)^2}$& $\displaystyle \frac{1}{2} + \frac{\omega_{32}\,\Omega_1}{8\, \mu_{23}^2}$ & 0 \\[5mm]
$\bra \op{A}_{33}\ket$ &0&$\displaystyle \frac{1}{2} - \frac{\omega_{31}\,\Omega_1}{8\, (\mu_{13}^2+\mu_{23}^2)}$& $\displaystyle \frac{1}{2} - \frac{\omega_{32}\,\Omega_1}{8\, \mu_{23}^2}$ & $\displaystyle \frac{1}{2} + \frac{\omega_{43}\,\Omega_2}{8\, \mu_{34}^2}$ \\[5mm]
$\bra \op{A}_{44}\ket$ &0&0&0&  $\displaystyle \frac{1}{2} - \frac{\omega_{43}\,\Omega_2}{8\, \mu_{34}^2}$ \\[2mm]
\end{tabular}
\end{table*}

\subsection{Quantum ground state}

Let us define the operator $\op{\Pi}(\op{K})=\exp(i\,\pi\,\op{K})$.  Imposing $\left[\op{\Pi}(\op{K}),\op{H}_{\lambda}\right]=0$ we obtain $3$ constants of motion of the system associated to
\begin{eqnarray}
\op{K}_{1} &=& \op{A}_{11} + \op{A}_{22} -\op{\nu}_1\,,\\[3mm]
\op{K}_{2} &=& \op{A}_{33} + \op{\nu}_{1} -\op{\nu}_2\,,\\[3mm]
\op{K}_{3} &=& \op{A}_{44}  +\op{\nu}_2\,.
\end{eqnarray}
These include the total number of atoms $\op{K}_{1}+\op{K}_{2}+\op{K}_{3} = N_a\,\op{I}$, as well as the total number of excitations $\op{M}  = 2 \op{K}_3 + \op{K}_2 = \op{\nu}_1+\op{\nu}_2 + \op{A}_{33} + 2\,\op{A}_{44}$. Note that $\op{K}_{3}$ corresponds to the total number of excitations of the mode $\Omega_2$. 

The basis of the problem is denoted by $|\nu_1\,\nu_2\,n_1\,n_2\,n_3\,n_4\ket$; using the operators $\op{M},\, \op{K}_{3}$, and the fact that $N_a = n_1+n_2+n_3+n_4$, it may be rewritten as 
\begin{displaymath}
|M-K_{3}-n_3-n_4,\,K_{3}-n_4,\,N_a-n_2-n_3-n_4,\,n_2,\,n_3,\,n_4\ket\,.
\end{displaymath}

The Hilbert space $\mathcal{H}$ of the system is thus divided into four subspaces, for the even and odd values of $M$ and $K_{3}$ 
\begin{center}
\begin{tabular}{c | c c}
subspace & $M$ & $K_{3}$ \\ \hline && \\[-2mm]
$\mathcal{H}_{ee} = \mathscr{L}\{|\phi\ket_{ee}\}$ & even & even \\[1mm]
$\mathcal{H}_{eo} = \mathscr{L}\{|\phi\ket_{eo}\}$ & even & odd \\[1mm]
$\mathcal{H}_{oe} = \mathscr{L}\{|\phi\ket_{oe}\}$ & odd & even \\[1mm]
$\mathcal{H}_{oo} = \mathscr{L}\{|\phi\ket_{oo}\}$ & odd & odd 
\end{tabular}
\end{center}
with minimum energies $E_{ee},\,E_{eo},\,E_{oe}$ and $E_{oo}$ respectively, and hence the ground quantum energy value is given by
\begin{equation}
E_g = \min\{E_{ee},\,E_{eo},\,E_{oe},\,E_{oo}\}\,.
\label{minEg}
\end{equation} 

The calculation is performed by truncating the basis at a value $M_{max}$ of $M$ which guarantees convergence of the lowest eigenvalue $E$ in each Hilbert subspace, and via Eq.~(\ref{minEg}) we determine the ground state eigenvalue $E_g$ and its corresponding eigenvector $\vert\phi_g\ket$. 

\begin{figure}[!h]
\includegraphics[width=0.85\linewidth]{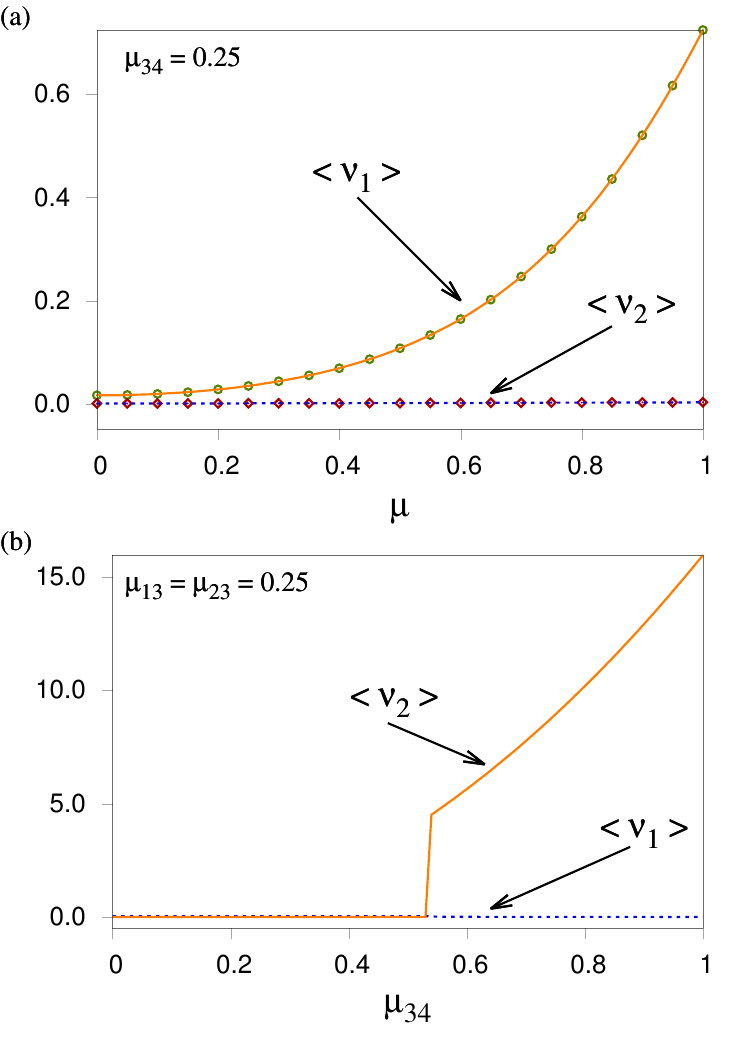}
\caption{(color online) Expectation values of the number of photons $\bra\op{\nu}_1\ket$ and $\bra \op{\nu}_2\ket$ for the $\lambda$-configuration. (a) $\bra\op{\nu}_1\ket$ (empty circles) and $\bra\op{\nu}_2\ket$ (empty diamonds) are plotted as functions of $\mu=\mu_{23}$, for $\mu_{13}=\mu_{34}=0.25$; and $\bra\op{\nu}_1\ket$ (continuous line) and $\bra\op{\nu}_2\ket$ (dashed line) are plotted as functions of $\mu=\mu_{13}$, for $\mu_{23}=\mu_{34}=0.25$. We see that the bulk of the quantum ground state remains in a subspace which only has contribution from the $\Omega_1$-mode. (b) $\bra\op{\nu}_1\ket$ (dashed line) and $\bra\op{\nu}_2\ket$ (continuous line) for $\mu_{13}=\mu_{23}=0.25$ as functions of $\mu_{34}$. In this region the bulk of the ground state is in a subspace which only has contribution from mode $\Omega_2$. All other parameters are as in Fig.~\ref{f.sep.l}. }\label{f.lambdaQ}
\end{figure}

In order to exemplify we consider the case $N_a=1$ and evaluate for this ground state the expectation values of the number of photons $\bra\op{\nu}_1\ket$ and $\bra\op{\nu}_2\ket$. We confirm the fact that the collective region is divided in monochromatic regions (cf. Fig.~\ref{f.sep.l}), and obtain the results shown in Figure~\ref{f.lambdaQ}. In sub-figure (a) $\bra\op{\nu}_1\ket$ (empty circles) and $\bra\op{\nu}_2\ket$ (empty diamonds) are plotted as functions of $\mu=\mu_{23}$, for $\mu_{13}=\mu_{34}=0.25$; and $\bra\op{\nu}_1\ket$ (continuous line) and $\bra\op{\nu}_2\ket$ (dashed line) are plotted as functions of $\mu=\mu_{13}$, for $\mu_{23}=\mu_{34}=0.25$. We see that the bulk of the quantum ground state remains in a subspace which only has contribution from the $\Omega_1$-mode. Sub-figure (b) shows $\bra\op{\nu}_1\ket$ (dashed line) and $\bra\op{\nu}_2\ket$ (continuous line) for $\mu_{13}=\mu_{23}=0.25$ as functions of $\mu_{34}$. In this region the bulk of the ground state is in a subspace which only has contribution from mode $\Omega_2$.

\section{$N$-configuration}
\label{s.N}

As a second example we consider $4$-level atoms in the $N$-configuration interacting with two modes of electromagnetic field. We consider the case when the mode $\Omega_1$ promotes transitions between levels $1\rightleftharpoons 3$ and $2\rightleftharpoons 4$, and the mode $\Omega_2$ promotes transitions between levels $2\rightleftharpoons 3$ (cf. Fig.~\ref{f.examples}). 
%
\begin{figure}[!h]
\includegraphics[width=0.95\linewidth]{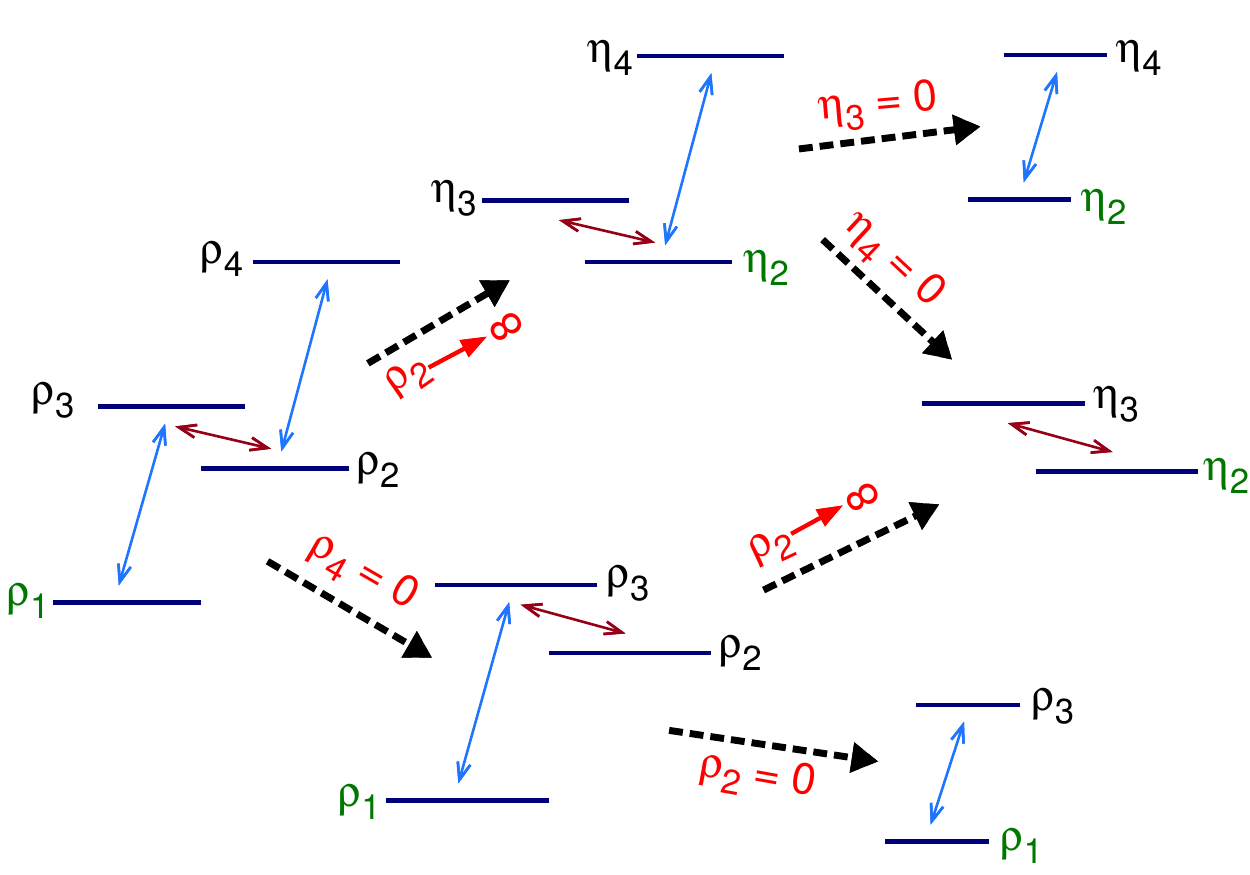}
\caption{(color online). Reductions of the $N$-configuration to $2$-level subsystems that allows to obtain the minimum energy surface.}\label{N_red}
\end{figure}
%
\begin{figure}[!h]
\includegraphics[width=0.85\linewidth]{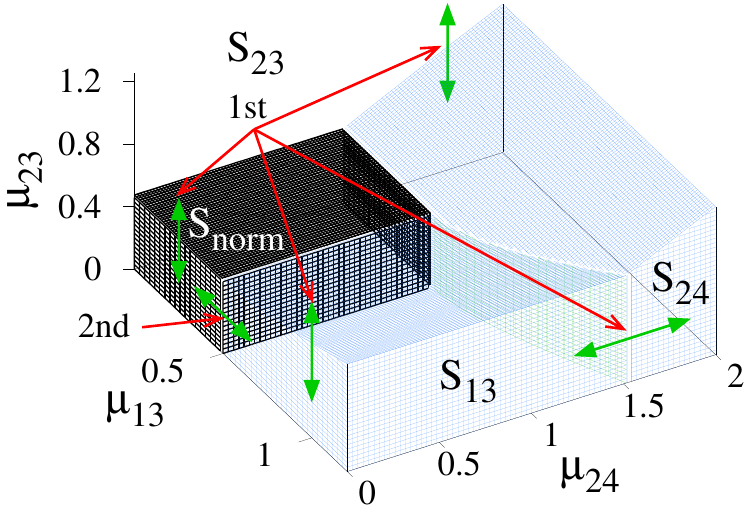}
\caption{(color online). Phase diagram for the $4$-level system in the $N$-configuration interacting dipolarly with two modes of electromagnetic field (cf.~Fig.~\ref{f.examples}). The normal region is shown in in black, and the collective region is divided by a separatrix (blue surface), below which the mode $\Omega_1$ contributes to the atomic transitions, and above which the mode $\Omega_2$ contributes. Further, the region where the mode $\Omega_1$ dominates is itself divided by a separatrix (green surface) which determines which of the $2$ subsystems is excited. The order of each transition is indicated. The parameters are $\Omega_1=1,\,\Omega_2=0.25,\,\omega_1=0,\, \omega_2=0.8,\,\omega_3= 1$ and $\omega_4=1.9$.}\label{f.sepN}
\end{figure}

\subsection{Variational ground state}

The critical values for the phases $\theta_i,\,\phi_{ij}$ take the form~(\ref{crit_ang}) while those for $r_i^c$ are given by
\begin{eqnarray}
	\label{crit_rN}
r_1^c &=& \frac{2\,\left(\mu_{13}\,\varrho_3^c+\mu_{24}\,\varrho_2^c\,\varrho_4^c\right)}{\Omega_1\,\Gamma_0^{c\,2}} \,,\quad
r_2^c = \frac{2\,\, \mu_{23}\,\varrho_2^c\, \varrho_3^c}{\Omega_2\,\Gamma_0^{c\,2}} \,.\qquad 
\end{eqnarray}
Following a similar procedure as in the previous example one finds four analytical expressions for the minimum energy surface: 
\begin{eqnarray}
E_{norm} &=& \omega_1\,,\\[3mm]
E_{13} &=& \omega_1 - \frac{(4\,\mu_{13}^2-(\omega_3-\omega_1)\Omega_1)^2}{16\,\mu_{13}^2\,\Omega_1}\,,\\[3mm]
E_{23} &=& \omega_2 - \frac{(4\,\mu_{23}^2-(\omega_3-\omega_2)\Omega_2)^2}{16\,\mu_{23}^2\,\Omega_2}\,,\\[3mm]
E_{24} &=& \omega_2 - \frac{(4\,\mu_{24}^2-(\omega_4-\omega_2)\Omega_1)^2}{16\,\mu_{24}^2\,\Omega_1}\,,
\end{eqnarray} 
where the expressions for the different $E_{ij}$ are valid when $4\,\mu_{ij}^2-(\omega_j-\omega_i)\Omega'\geq0$ is satisfied. In this expression $\Omega'$ stands for the corresponding mode. The minimum energy surface is given by
\begin{equation}
E_{v} = \min\{E_N,\,E_{13},\,E_{23},\,E_{24}\}\,.
\end{equation} 

The expressions for the energy are associated with the reduction shown in Fig.~\ref{N_red}.
$E_{norm}$ corresponds to $\varrho_{i}^{c}=0$, $i=2,\,3,\,4$. $E_{13}$ is obtained by taking successively $\varrho_4=0$ and $\varrho_2=0$, arriving to $2$-levels.  Similarly, for $E_{24}$ we take $\varrho_2\to\infty$ and $\eta_3=0$, while $E_{23}$ may be obtained following two paths, viz., $\varrho_4=0$ and $\varrho_2\to\infty$ or $\varrho_2\to\infty$ and $\eta_4=0$.

The separatrix defined by the set of values where $E_{norm} = E_{ij}$ and $E_{ij}=E_{lm}$ is shown in figure~\ref{f.sepN}. The normal region with $E_{v}=E_{norm}$, the collective regions where $E_{v}=E_{ij}$, and the order of transitions, are indicated. One may observe that a second order transition is had only for $S_{norm}\rightleftharpoons S_{13}$, and a first order transition for other cases. Once again, the collective region is divided into monochromatic subregions, and the region where the mode $\Omega_1$ dominates is divided into two subregions $S_{13}$ and $S_{24}$, each one corresponding to different subsystems of two levels.

The critical values of the Hamiltonian variables in each region are given in Table~\ref{t.N}, and 
the relevant expectation values of the field and matter observables calculated from the variational states analytically are displayed in Table~\ref{t.N.ev}.

\begin{table*}[!ht]
\caption{Critical values for the $N$-configuration, we chose $\omega_1=0$}\label{t.N}
\vspace{2mm}
\begin{tabular}{l | c c c c c}
& $r_1^c$& $r_2^c$& $\varrho_2^c$ & $\varrho_3^c$ & $\varrho_4^c$ \\[1mm] \hline \hline  \\
$S_{norm}$ &0&0& 0&0&0 \\[2mm]
$S_{13}$ & $\displaystyle \frac{\sqrt{16 \mu_{13}^4-\omega_{31}^2\,\Omega_1^2}}{4\,\mu_{13}\,\Omega_1}$& 0 & $0 $ & $\displaystyle \sqrt{\frac{4\,\mu_{13}^2- \omega_{31}\,\Omega_1}{4\,\mu_{13}^2+ \omega_{31}\,\Omega_1}}$ & 0 \\[2mm]
&&&&& \\
\hline&& &&& \\[-3mm] &&& $\varrho_2^c\to \infty$ &$\eta_3^c$& $\eta_4^c$\\[1mm]
\hline &&&&& \\[-2mm]
$S_{24}$ & $\displaystyle \frac{\sqrt{16 \mu_{24}^4 - \omega_{42}^2\,\Omega_1^2}}{4\,\mu_{24}\,\Omega_1}$&0 & & 0 & $\displaystyle \sqrt{\frac{4\,\mu_{24}^2 - \omega_{42}\,\Omega_1}{4\,\mu_{24}^2 + \omega_{42}\,\Omega_1}}$\\[2mm]
$S_{23}$&0&$\displaystyle \frac{\sqrt{16 \mu_{23}^4-\omega_{32}^2\,\Omega_2^2}}{4\,\mu_{23}\,\Omega_2}$&&$\displaystyle \sqrt{\frac{4\,\mu_{23}^2- \omega_{32}\,\Omega_2}{4\,\mu_{23}^2+ \omega_{32}\,\Omega_2}}$&0\\[3mm]
\end{tabular}
\end{table*}

\begin{table*}
\caption{Expectation values per particle in the different regions, for the $N$-configuration.}\label{t.N.ev}
\vspace{2mm}
\begin{tabular}{c| c | c | c | c}
& $S_{norm}$& $S_{13}$& $S_{23}$ & $S_{24}$   \\[1mm] \hline \hline  &&&& \\
$\bra \op{H}_N\ket$ &0&$\omega_1-\displaystyle \frac{(4\,\mu_{13}^2- \omega_{31}\,\Omega_1)^2}{16\,\mu_{13}^2\,\Omega_1}$& $\displaystyle \omega_2 - \frac{(4\,\mu_{23}^2-\omega_{32}\,\Omega_2)^2}{16\,\mu_{23}^2\,\Omega_2}$ &$\displaystyle \omega_2 - \frac{(4\,\mu_{24}^2-\omega_{42}\,\Omega_1)^2}{16\,\mu_{24}^2\,\Omega_1}$\\[5mm]
$\bra \op{\nu}_1\ket$&0 &$\displaystyle \frac{16\, \mu_{13}^4 - \omega_{31}^2\,\Omega_1^2}{16\,\mu_{13}^2\,\Omega_1^2}$& 0 &$\displaystyle \frac{16\, \mu_{24}^4 - \omega_{42}^2\,\Omega_1^2}{16\,\mu_{24}^2\,\Omega_1^2}$ \\[5mm]
$\bra \op{\nu}_2\ket$ &0&0& $\displaystyle \frac{16\, \mu_{23}^4 - \omega_{32}^2\,\Omega_2^2}{16\,\mu_{23}^2\,\Omega_2^2}$  &0 \\[5mm]
$\bra \op{A}_{11}\ket$ &1&$\displaystyle \frac{4\,\mu_{13}^2 + \omega_{31}\,\Omega_1}{8\,\mu_{13}^2}$ & 0 &0\\[5mm]
$\bra \op{A}_{22}\ket$ &0&0& $\displaystyle \frac{4\,\mu_{23}^2 + \omega_{32}\,\Omega_2}{8\,\mu_{23}^2}$  &$\displaystyle \frac{4\,\mu_{24}^2 + \omega_{42}\,\Omega_1}{8\,\mu_{24}^2}$ \\[5mm]
$\bra \op{A}_{33}\ket$ &0&$\displaystyle \frac{4\,\mu_{13}^2 - \omega_{31}\,\Omega_1}{8\,\mu_{13}^2}$& $\displaystyle \frac{4\,\mu_{23}^2 - \omega_{32}\,\Omega_2}{8\,\mu_{23}^2}$ &0\\[5mm]
$\bra \op{A}_{44}\ket$ &0&0& 0 &$\displaystyle \frac{4\,\mu_{24}^2 - \omega_{42}\,\Omega_1}{8\,\mu_{24}^2}$\\[5mm]
\end{tabular}
\end{table*}

\subsection{Quantum ground state}

In the case of the $N$-configuration, besides $N_{a}$ the constant of motion is the parity of the total number of excitations
$\op{\Pi}(\op{M})=\exp(i\,\pi\,\op{M})$, with
\begin{eqnarray}
\op{M}=\op{A}_{33}+\op{A}_{44} +\op{\nu}_1+\op{\nu}_2\,.
\end{eqnarray}
On may then define subspaces which preserve the parity of $M$ (even or odd) and find the minimum energy values $E_e$ and $E_o$, respectively. The quantum ground energy will then given by 
\begin{equation}
E_g = \min\left\{E_e,\,E_o\right\}\,.
\end{equation} 

For $N_a=1$ the ground state possesses an even parity of $M$, i.e, $E_g=E_e$.  As in the variational case (cf. Fig.~\ref{f.sepN}), the mode $\Omega_1$ dominates ($\bra\op{\nu}_1\ket\gg\bra\op{\nu}_2\ket$) in the regions $S_{13}$ and $S_{24}$ while the mode $\Omega_2$ dominates ($\bra\op{\nu}_1\ket\ll\bra\op{\nu}_2\ket$) in the region $S_{23}$.

\begin{figure}[!ht]
\includegraphics[width=0.85\linewidth]{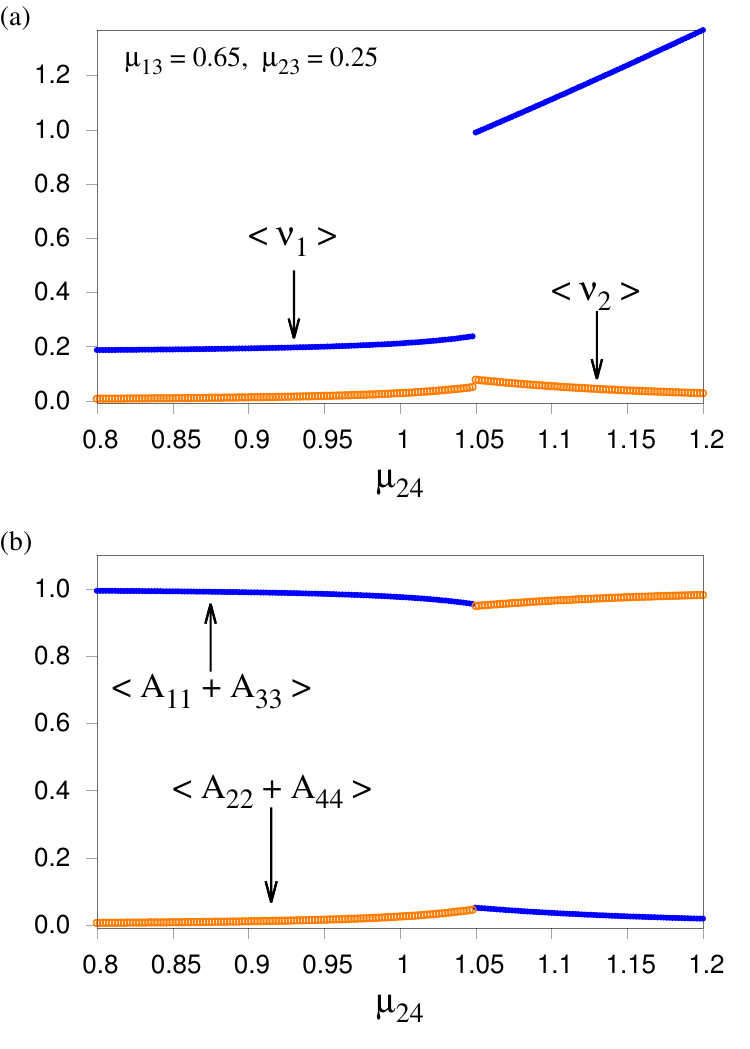}
\caption{(color online). (a) Expectation values of the number of photons $\bra\op{\nu}_1\ket$ (full circles) and $\bra \op{\nu}_2\ket$ (empty circles) for the $N$-configuration. (b) Atomic populations $\bra \op{A}_{11}+\op{A}_{33}\ket$ (full circles) and  $\bra \op{A}_{22}+\op{A}_{44}\ket$ (empty circles) corresponding to the two $2$-level subsystems of the mode $\Omega_1$. Parameters are the same that in Fig.~\ref{f.sepN}. See text.}\label{f.sepNQ}
\end{figure}

Focusing our attention on the transition $S_{13}\rightleftharpoons S_{24}$, where the mode $\Omega_1$ dominates, and using the same parameters as for the variational case, $\mu_{13}=0.65$ and $\mu_{23}=0.25$, we evaluate the expectation value of the number of photons $\bra \op{\nu}_1\ket$ and $\bra\op{\nu}_2\ket$ as functions of $\mu_{24}$. Fig.~\ref{f.sepNQ}(a) shows a discontinuity in the expectation value of both $\bra\op{\nu}_1\ket$ and $\bra\op{\nu}_2\ket$ at the transition, due to the fact that the critical points in the separatrix form a Maxwell set. We also calculate the atomic populations for the two $2$-level subsystems, $\bra \op{A}_{11}+\op{A}_{33}\ket$ and $\bra \op{A}_{22}+\op{A}_{44}\ket$, shown in Fig.~\ref{f.sepNQ}(b). In the region $S_{13}$ we have $\bra \op{A}_{11}+\op{A}_{33}\ket\approx N_a\,(=1)$, i.e., the quantum ground state is practically the state of the single subsystem of $2$-levels formed by the first and third atomic levels, while in region $S_{24}$ it corresponds to the subsystem formed by the second and fourth atomic levels $\bra \op{A}_{22}+\op{A}_{44}\ket\approx N_a$. In other words, the transition in this region is due to the change of excitations in the atomic subsystem, in accordance with the variational calculation.

%

\section{Conclusions}
\label{conclusions}

Two $4$-level atomic configurations under dipolar interaction with $2$ modes of electromagnetic radiation, viz., the $\lambda$- and $N$ configurations, have been shown to present qualitatively different quantum phase diagrams. Both configurations exhibit normal and collective (super-radiant) regimes; while the latter in the $\lambda$-configuration divides itself into $2$ subregions, corresponding to each of the modes, that in the $N$-configuration may be divided into $2$ or $3$ subregions depending on whether the field modes divide the atomic system into $2$ separate subsystems or not. The quantum phase diagrams are obtained by means of a variational procedure, which allows for analytical expressions both for the critical variables and the expectation values of the field and matter operators, and confirmed through the exact quantum calculation (even for a value of $N_a$ as low as $1$).

The parameter space is divided into monochromatic subregions $S_{ij}$, each of which is dominated by a mode of the radiation field, in agreement with~\cite{cordero15}.  
In general, both first and second order transitions for $S_{norm}~\rightleftharpoons~S_{ij}$ occur~\cite{cordero2}. A first order transition occurs when the critical points along the separatrix form a Maxwell set~\cite{castanos14,cordero15}; a second-order transition is related to the fact that the critical points form bifurcations.

It is interesting to note that for first order transitions we have discontinuities in the expectation values of the photon number and atomic population operators. This is also true for the $N$-configuration for the transition $S_{13}~ \rightleftharpoons~S_{24}$ in spite of the fact that we have only one mode. This is as consequence of the atomic population shift from one $2$-level subsystem to another.

In order to recover quantum correlations we need to restore the symmetries of the corresponding Hamiltonians by acting with the parity operators on the variational state, as it is usually done in many-body theories.

\

\section*{Acknowledgements}
\noindent
This work was partially supported by CONACyT-M\'exico (under Project No.~238494), and DGAPA-UNAM (under Projects No.~IN101614 and No.~IN110114).


\begin{thebibliography}{19}%
\makeatletter
\providecommand \@ifxundefined [1]{%
 \@ifx{#1\undefined}
}%
\providecommand \@ifnum [1]{%
 \ifnum #1\expandafter \@firstoftwo
 \else \expandafter \@secondoftwo
 \fi
}%
\providecommand \@ifx [1]{%
 \ifx #1\expandafter \@firstoftwo
 \else \expandafter \@secondoftwo
 \fi
}%
\providecommand \natexlab [1]{#1}%
\providecommand \enquote  [1]{``#1''}%
\providecommand \bibnamefont  [1]{#1}%
\providecommand \bibfnamefont [1]{#1}%
\providecommand \citenamefont [1]{#1}%
\providecommand \href@noop [0]{\@secondoftwo}%
\providecommand \href [0]{\begingroup \@sanitize@url \@href}%
\providecommand \@href[1]{\@@startlink{#1}\@@href}%
\providecommand \@@href[1]{\endgroup#1\@@endlink}%
\providecommand \@sanitize@url [0]{\catcode `\\12\catcode `\$12\catcode
  `\&12\catcode `\#12\catcode `\^12\catcode `\_12\catcode `\%12\relax}%
\providecommand \@@startlink[1]{}%
\providecommand \@@endlink[0]{}%
\providecommand \url  [0]{\begingroup\@sanitize@url \@url }%
\providecommand \@url [1]{\endgroup\@href {#1}{\urlprefix }}%
\providecommand \urlprefix  [0]{URL }%
\providecommand \Eprint [0]{\href }%
\providecommand \doibase [0]{http://dx.doi.org/}%
\providecommand \selectlanguage [0]{\@gobble}%
\providecommand \bibinfo  [0]{\@secondoftwo}%
\providecommand \bibfield  [0]{\@secondoftwo}%
\providecommand \translation [1]{[#1]}%
\providecommand \BibitemOpen [0]{}%
\providecommand \bibitemStop [0]{}%
\providecommand \bibitemNoStop [0]{.\EOS\space}%
\providecommand \EOS [0]{\spacefactor3000\relax}%
\providecommand \BibitemShut  [1]{\csname bibitem#1\endcsname}%
\let\auto@bib@innerbib\@empty
\bibitem [{\citenamefont {Dicke}(1954)}]{dicke54}%
  \BibitemOpen
  \bibfield  {author} {\bibinfo {author} {\bibfnamefont {R.~H.}\ \bibnamefont
  {Dicke}},\ }\href {\doibase 10.1103/PhysRev.93.99} {\bibfield  {journal}
  {\bibinfo  {journal} {Phys. Rev.}\ }\textbf {\bibinfo {volume} {93}},\
  \bibinfo {pages} {99} (\bibinfo {year} {1954})}\BibitemShut {NoStop}%
\bibitem [{\citenamefont {Hepp}\ and\ \citenamefont
  {Lieb}(1973{\natexlab{a}})}]{hepp73}%
  \BibitemOpen
  \bibfield  {author} {\bibinfo {author} {\bibfnamefont {K.}~\bibnamefont
  {Hepp}}\ and\ \bibinfo {author} {\bibfnamefont {E.~H.}\ \bibnamefont
  {Lieb}},\ }\href {\doibase 10.1016/0003-4916(73)90039-0} {\bibfield
  {journal} {\bibinfo  {journal} {Annals of Physics}\ }\textbf {\bibinfo
  {volume} {76}},\ \bibinfo {pages} {360} (\bibinfo {year}
  {1973}{\natexlab{a}})}\BibitemShut {NoStop}%
\bibitem [{\citenamefont {Hepp}\ and\ \citenamefont
  {Lieb}(1973{\natexlab{b}})}]{hepp73b}%
  \BibitemOpen
  \bibfield  {author} {\bibinfo {author} {\bibfnamefont {K.}~\bibnamefont
  {Hepp}}\ and\ \bibinfo {author} {\bibfnamefont {E.~H.}\ \bibnamefont
  {Lieb}},\ }\href {\doibase 10.1103/PhysRevA.8.2517} {\bibfield  {journal}
  {\bibinfo  {journal} {Phys. Rev. A}\ }\textbf {\bibinfo {volume} {8}},\
  \bibinfo {pages} {2517} (\bibinfo {year} {1973}{\natexlab{b}})}\BibitemShut
  {NoStop}%
\bibitem [{\citenamefont {Kastner}(1993)}]{kastner93}%
  \BibitemOpen
  \bibfield  {author} {\bibinfo {author} {\bibfnamefont {M.~A.}\ \bibnamefont
  {Kastner}},\ }\href@noop {} {\bibfield  {journal} {\bibinfo  {journal} {Phys.
  Today}\ }\textbf {\bibinfo {volume} {46}},\ \bibinfo {pages} {24} (\bibinfo
  {year} {1993})}\BibitemShut {NoStop}%
\bibitem [{\citenamefont {Astafiev}\ \emph {et~al.}(2010)\citenamefont
  {Astafiev}, \citenamefont {Zagoskin}, \citenamefont {{Abdumalikov Jr.}},
  \citenamefont {Pashkin}, \citenamefont {Yamamoto}, \citenamefont {Inomata},
  \citenamefont {Nakamura},\ and\ \citenamefont {Tsai}}]{astafiev10}%
  \BibitemOpen
  \bibfield  {author} {\bibinfo {author} {\bibfnamefont {O.}~\bibnamefont
  {Astafiev}}, \bibinfo {author} {\bibfnamefont {A.~M.}\ \bibnamefont
  {Zagoskin}}, \bibinfo {author} {\bibfnamefont {A.~A.}\ \bibnamefont
  {{Abdumalikov Jr.}}}, \bibinfo {author} {\bibfnamefont {Y.~A.}\ \bibnamefont
  {Pashkin}}, \bibinfo {author} {\bibfnamefont {T.}~\bibnamefont {Yamamoto}},
  \bibinfo {author} {\bibfnamefont {K.}~\bibnamefont {Inomata}}, \bibinfo
  {author} {\bibfnamefont {Y.}~\bibnamefont {Nakamura}}, \ and\ \bibinfo
  {author} {\bibfnamefont {J.~S.}\ \bibnamefont {Tsai}},\ }\href {\doibase
  10.1126/science.1181918} {\bibfield  {journal} {\bibinfo  {journal}
  {Science}\ }\textbf {\bibinfo {volume} {327}},\ \bibinfo {pages} {840}
  (\bibinfo {year} {2010})}\BibitemShut {NoStop}%
\bibitem [{\citenamefont {Buluta}\ \emph {et~al.}(2011)\citenamefont {Buluta},
  \citenamefont {Ashhab},\ and\ \citenamefont {Nori}}]{buluta11}%
  \BibitemOpen
  \bibfield  {author} {\bibinfo {author} {\bibfnamefont {I.}~\bibnamefont
  {Buluta}}, \bibinfo {author} {\bibfnamefont {S.}~\bibnamefont {Ashhab}}, \
  and\ \bibinfo {author} {\bibfnamefont {F.}~\bibnamefont {Nori}},\ }\href
  {\doibase 10.1088/0034-4885/74/10/104401} {\bibfield  {journal} {\bibinfo
  {journal} {Rep. Prog. Phys.}\ }\textbf {\bibinfo {volume} {74}},\ \bibinfo
  {pages} {104401} (\bibinfo {year} {2011})}\BibitemShut {NoStop}%
\bibitem [{\citenamefont {Baumann}\ \emph {et~al.}(2010)\citenamefont
  {Baumann}, \citenamefont {Guerlin}, \citenamefont {Brennecke},\ and\
  \citenamefont {Esslinger}}]{baumann10}%
  \BibitemOpen
  \bibfield  {author} {\bibinfo {author} {\bibfnamefont {K.}~\bibnamefont
  {Baumann}}, \bibinfo {author} {\bibfnamefont {C.}~\bibnamefont {Guerlin}},
  \bibinfo {author} {\bibfnamefont {F.}~\bibnamefont {Brennecke}}, \ and\
  \bibinfo {author} {\bibfnamefont {T.}~\bibnamefont {Esslinger}},\ }\href
  {\doibase doi:10.1038/nature09009} {\bibfield  {journal} {\bibinfo  {journal}
  {Nature}\ }\textbf {\bibinfo {volume} {464}},\ \bibinfo {pages} {1301–1306}
  (\bibinfo {year} {2010})}\BibitemShut {NoStop}%
\bibitem [{\citenamefont {Baksic}\ \emph {et~al.}(2013)\citenamefont {Baksic},
  \citenamefont {Nataf},\ and\ \citenamefont {Ciuti}}]{baksic13}%
  \BibitemOpen
  \bibfield  {author} {\bibinfo {author} {\bibfnamefont {A.}~\bibnamefont
  {Baksic}}, \bibinfo {author} {\bibfnamefont {P.}~\bibnamefont {Nataf}}, \
  and\ \bibinfo {author} {\bibfnamefont {C.}~\bibnamefont {Ciuti}},\ }\href
  {\doibase 10.1103/PhysRevA.87.023813} {\bibfield  {journal} {\bibinfo
  {journal} {Phys. Rev. A}\ }\textbf {\bibinfo {volume} {87}},\ \bibinfo
  {pages} {023813} (\bibinfo {year} {2013})}\BibitemShut {NoStop}%
\bibitem [{\citenamefont {Yoo}\ and\ \citenamefont {Eberly}(1985)}]{yoo85}%
  \BibitemOpen
  \bibfield  {author} {\bibinfo {author} {\bibfnamefont {H.~I.}\ \bibnamefont
  {Yoo}}\ and\ \bibinfo {author} {\bibfnamefont {J.~H.}\ \bibnamefont
  {Eberly}},\ }\href {\doibase 10.1016/0370-1573(85)90015-8} {\bibfield
  {journal} {\bibinfo  {journal} {Phys. Rep.}\ }\textbf {\bibinfo {volume}
  {118}},\ \bibinfo {pages} {239} (\bibinfo {year} {1985})}\BibitemShut
  {NoStop}%
\bibitem [{\citenamefont {Abdel-Wahab}(2007)}]{abdel-wahab07}%
  \BibitemOpen
  \bibfield  {author} {\bibinfo {author} {\bibfnamefont {N.~H.}\ \bibnamefont
  {Abdel-Wahab}},\ }\href {\doibase 10.1088/0031-8949/76/3/006} {\bibfield
  {journal} {\bibinfo  {journal} {Phys. Scr.}\ }\textbf {\bibinfo {volume}
  {76}},\ \bibinfo {pages} {244} (\bibinfo {year} {2007})}\BibitemShut
  {NoStop}%
\bibitem [{\citenamefont {Abdel-Wahab}(2008)}]{abdel-wahab08}%
  \BibitemOpen
  \bibfield  {author} {\bibinfo {author} {\bibfnamefont {N.~H.}\ \bibnamefont
  {Abdel-Wahab}},\ }\href {\doibase 10.1142/S0217984908016868} {\bibfield
  {journal} {\bibinfo  {journal} {Mod. Phys. Lett. B}\ }\textbf {\bibinfo
  {volume} {22}},\ \bibinfo {pages} {2587} (\bibinfo {year}
  {2008})}\BibitemShut {NoStop}%
\bibitem [{\citenamefont {Civitarese}\ and\ \citenamefont
  {Reboiro}(2006)}]{civitarese1}%
  \BibitemOpen
  \bibfield  {author} {\bibinfo {author} {\bibfnamefont {O.}~\bibnamefont
  {Civitarese}}\ and\ \bibinfo {author} {\bibfnamefont {M.}~\bibnamefont
  {Reboiro}},\ }\href {\doibase 10.1016/j.physleta.2006.04.043} {\bibfield
  {journal} {\bibinfo  {journal} {Physics Letters A}\ }\textbf {\bibinfo
  {volume} {357}},\ \bibinfo {pages} {224 } (\bibinfo {year}
  {2006})}\BibitemShut {NoStop}%
\bibitem [{\citenamefont {Cordero}\ \emph
  {et~al.}(2013{\natexlab{a}})\citenamefont {Cordero}, \citenamefont
  {{L\'opez-Pe\~na}}, \citenamefont {{Casta\~nos}},\ and\ \citenamefont
  {{Nahmad-Achar}}}]{cordero1}%
  \BibitemOpen
  \bibfield  {author} {\bibinfo {author} {\bibfnamefont {S.}~\bibnamefont
  {Cordero}}, \bibinfo {author} {\bibfnamefont {R.}~\bibnamefont
  {{L\'opez-Pe\~na}}}, \bibinfo {author} {\bibfnamefont {O.}~\bibnamefont
  {{Casta\~nos}}}, \ and\ \bibinfo {author} {\bibfnamefont {E.}~\bibnamefont
  {{Nahmad-Achar}}},\ }\href {\doibase 10.1103/PhysRevA.87.023805} {\bibfield
  {journal} {\bibinfo  {journal} {Phys. Rev. A}\ }\textbf {\bibinfo {volume}
  {87}},\ \bibinfo {pages} {023805} (\bibinfo {year}
  {2013}{\natexlab{a}})}\BibitemShut {NoStop}%
\bibitem [{\citenamefont {Cordero}\ \emph
  {et~al.}(2013{\natexlab{b}})\citenamefont {Cordero}, \citenamefont
  {{Casta\~nos}}, \citenamefont {{L\'opez-Pe\~na}},\ and\ \citenamefont
  {{Nahmad-Achar}}}]{cordero2}%
  \BibitemOpen
  \bibfield  {author} {\bibinfo {author} {\bibfnamefont {S.}~\bibnamefont
  {Cordero}}, \bibinfo {author} {\bibfnamefont {O.}~\bibnamefont
  {{Casta\~nos}}}, \bibinfo {author} {\bibfnamefont {R.}~\bibnamefont
  {{L\'opez-Pe\~na}}}, \ and\ \bibinfo {author} {\bibfnamefont
  {E.}~\bibnamefont {{Nahmad-Achar}}},\ }\href {\doibase
  10.1088/1751-8113/46/50/505302} {\bibfield  {journal} {\bibinfo  {journal}
  {J. Phys. A: Math. Theor.}\ }\textbf {\bibinfo {volume} {46}},\ \bibinfo
  {pages} {505302} (\bibinfo {year} {2013}{\natexlab{b}})}\BibitemShut
  {NoStop}%
\bibitem [{\citenamefont {Hayn}\ \emph {et~al.}(2011)\citenamefont {Hayn},
  \citenamefont {Emary},\ and\ \citenamefont {Brandes}}]{hayn11}%
  \BibitemOpen
  \bibfield  {author} {\bibinfo {author} {\bibfnamefont {M.}~\bibnamefont
  {Hayn}}, \bibinfo {author} {\bibfnamefont {C.}~\bibnamefont {Emary}}, \ and\
  \bibinfo {author} {\bibfnamefont {T.}~\bibnamefont {Brandes}},\ }\href
  {\doibase 10.1103/PhysRevA.84.053856} {\bibfield  {journal} {\bibinfo
  {journal} {Phys. Rev. A}\ }\textbf {\bibinfo {volume} {84}},\ \bibinfo
  {pages} {053856} (\bibinfo {year} {2011})}\BibitemShut {NoStop}%
\bibitem [{\citenamefont {Hayn}\ \emph {et~al.}(2012)\citenamefont {Hayn},
  \citenamefont {Emary},\ and\ \citenamefont {Brandes}}]{hayn12}%
  \BibitemOpen
  \bibfield  {author} {\bibinfo {author} {\bibfnamefont {M.}~\bibnamefont
  {Hayn}}, \bibinfo {author} {\bibfnamefont {C.}~\bibnamefont {Emary}}, \ and\
  \bibinfo {author} {\bibfnamefont {T.}~\bibnamefont {Brandes}},\ }\href
  {\doibase 10.1103/PhysRevA.86.063822} {\bibfield  {journal} {\bibinfo
  {journal} {Phys. Rev. A}\ }\textbf {\bibinfo {volume} {86}},\ \bibinfo
  {pages} {063822} (\bibinfo {year} {2012})}\BibitemShut {NoStop}%
\bibitem [{\citenamefont {Cordero}\ \emph {et~al.}(2015)\citenamefont
  {Cordero}, \citenamefont {Nahmad-Achar}, \citenamefont {L\'opez-Pe\~na},\
  and\ \citenamefont {Casta\~nos}}]{cordero15}%
  \BibitemOpen
  \bibfield  {author} {\bibinfo {author} {\bibfnamefont {S.}~\bibnamefont
  {Cordero}}, \bibinfo {author} {\bibfnamefont {E.}~\bibnamefont
  {Nahmad-Achar}}, \bibinfo {author} {\bibfnamefont {R.}~\bibnamefont
  {L\'opez-Pe\~na}}, \ and\ \bibinfo {author} {\bibfnamefont {O.}~\bibnamefont
  {Casta\~nos}},\ }\href {\doibase 10.1103/PhysRevA.92.053843} {\bibfield
  {journal} {\bibinfo  {journal} {Phys. Rev. A}\ }\textbf {\bibinfo {volume}
  {92}},\ \bibinfo {pages} {053843} (\bibinfo {year} {2015})}\BibitemShut
  {NoStop}%
\bibitem [{\citenamefont {L\'opez Pe\~na}\ \emph {et~al.}(2015)\citenamefont
  {L\'opez Pe\~na}, \citenamefont {Cordero}, \citenamefont {Nahmad-Achar},\
  and\ \citenamefont {{Casta\~nos}}}]{lopez-pena15}%
  \BibitemOpen
  \bibfield  {author} {\bibinfo {author} {\bibfnamefont {R.}~\bibnamefont
  {L\'opez Pe\~na}}, \bibinfo {author} {\bibfnamefont {S.}~\bibnamefont
  {Cordero}}, \bibinfo {author} {\bibfnamefont {E.}~\bibnamefont
  {Nahmad-Achar}}, \ and\ \bibinfo {author} {\bibfnamefont {O.}~\bibnamefont
  {{Casta\~nos}}},\ }\href {\doibase 10.1088/0031-8949/90/6/068016} {\bibfield
  {journal} {\bibinfo  {journal} {Phys. Scr.}\ }\textbf {\bibinfo {volume}
  {90}},\ \bibinfo {pages} {068016} (\bibinfo {year} {2015})}\BibitemShut
  {NoStop}%
\bibitem [{\citenamefont {{Casta\~nos}}\ \emph {et~al.}(2014)\citenamefont
  {{Casta\~nos}}, \citenamefont {Cordero}, \citenamefont {{L\'opez-Pe\~na}},\
  and\ \citenamefont {Nahmad-Achar}}]{castanos14}%
  \BibitemOpen
  \bibfield  {author} {\bibinfo {author} {\bibfnamefont {O.}~\bibnamefont
  {{Casta\~nos}}}, \bibinfo {author} {\bibfnamefont {S.}~\bibnamefont
  {Cordero}}, \bibinfo {author} {\bibfnamefont {R.}~\bibnamefont
  {{L\'opez-Pe\~na}}}, \ and\ \bibinfo {author} {\bibfnamefont
  {E.}~\bibnamefont {Nahmad-Achar}},\ }\href {\doibase
  10.1088/1742-6596/512/1/012006} {\bibfield  {journal} {\bibinfo  {journal}
  {J. Phys.: Conf. Ser.}\ }\textbf {\bibinfo {volume} {512}},\ \bibinfo {pages}
  {012006} (\bibinfo {year} {2014})}\BibitemShut {NoStop}%
\end{thebibliography}
\end{document}